\documentclass[conference]{IEEEtran}
\usepackage{cite}
\usepackage{graphicx}
\graphicspath{ {images/} }

\begin{document}


\title{Multiscale Granger causality analysis by {\it \`a trous} wavelet transform}


\author{\IEEEauthorblockN{Sebastiano Stramaglia\IEEEauthorrefmark{1}\IEEEauthorrefmark{2},
Iege Bassez\IEEEauthorrefmark{3},
Luca Faes\IEEEauthorrefmark{4} 
and Daniele Marinazzo\IEEEauthorrefmark{3}}
\IEEEauthorblockA{\IEEEauthorrefmark{1} Dipartimento di Fisica, Universita degli Studi ``Aldo Moro" Bari (Italy)}
\IEEEauthorblockA{\IEEEauthorrefmark{2} Istituto Nazionale di Fisica Nucleare Sezione di Bari (Italy)}
\IEEEauthorblockA{\IEEEauthorrefmark{3} Department of Data Analysis. Ghent University (Belgium) }
\IEEEauthorblockA{\IEEEauthorrefmark{4} BIOtech, Dept. of Industrial Engineering, University of Trento, and IRCS-PAT FBK, Trento (Italy)}
Email: sebastiano.stramaglia@ba.infn.it}


\maketitle


\begin{abstract}
Since interactions in neural systems occur across multiple temporal scales, it is likely that information flow will exhibit a multiscale structure, thus requiring a  multiscale generalization of classical temporal precedence causality analysis like Granger's approach. However, the computation  of  multiscale measures  of  information  dynamics  is  complicated  by theoretical  and  practical  issues such as filtering and undersampling: to overcome these problems, we propose a wavelet-based approach for multiscale Granger causality (GC) analysis, which is characterized by the following properties: (i) only the candidate driver variable is wavelet transformed (ii) the decomposition is performed using the {\it \`a trous} wavelet transform with cubic B-spline filter.  We measure GC, at a given scale,  by including the wavelet coefficients of the driver times series, at that scale,  in the regression model of the target. To validate our method, we apply it to publicly available scalp EEG signals, and we find that the condition of closed eyes, at rest, is characterized by an enhanced GC among channels at slow scales w.r.t. eye open condition, whilst the standard Granger causality is not significantly different in the two conditions.
\end{abstract}

\begin {IEEEkeywords}
Granger causality, multiscale analysis, Wavelet transform, scalp EEG 
\end{IEEEkeywords}

\section{Introduction}
\label{intro}
Great attention has been devoted in the last years to the identification of information flows in human brains. Wiener \cite{wiener} and Granger \cite{granger} formalized the
notion that, if the prediction of one time series could be improved
by incorporating the knowledge of past values of a second one, then
the latter is said to have a {\it causal} influence on the former.
Initially developed for econometric applications, Granger causality (GC)
has gained popularity also among engineers and physicists (see, e.g.,
\cite{book1,book2}). GC is connected to the information flow between
variables \cite{hla}. A kernel method
for GC, introduced in \cite{noiprl}, deals with the
nonlinear case by embedding data into a Hilbert space, and searching
for linear relations in that space. Geweke \cite{geweke} has
generalized GC to a multivariate fashion in order to
identify conditional GC; as described in
\cite{noipre}, multivariate GC may be used to infer the
structure of dynamical networks \cite{dn} from data.
Another important notion in information theory
is the redundancy in a group of variables, formalized in
\cite{palus} as a generalization of the mutual information. A
formalism to recognize redundant and synergetic variables in
dynamical networks has been proposed in \cite{stramagliaIEEE}; the information theoretic treatment of
groups of correlated degrees of freedom can reveal their functional
roles in complex systems.

-

The manuscript is organized as follows: in the next section we  describe the {\it \`a trous} wavelet transform, whilst in Section 3 we briefly recall the notion of GC and introduce the new method. In Section 4 we describe the application of the proposed approach to scalp EEG signals corresponding to resting conditions with closed eyes and with open eyes \cite{physionet}. Section 5 summarizes our conclusions.
\section{Wavelet transform}
\label{section2}
In the present section we give a brief account of discrete wavelet
mathematical aspects that are relevant to our objectives. The wavelet transform is a signal
processing technique that represents a transient or non-stationary
signal in terms of time and scale distribution, and it is an
excellent tool for on-line data compression, analysis and reducing,
etc. \cite{stark} The most striking difference between Fourier and wavelet
decomposition is that the last allows for a projection on modes
simultaneously localized in both time and frequency space, up to the
limit of classical uncertainty relations. Unlike the Fourier bases,
which are delocalized by definition, the wavelet bases have compact
spatial support, therefore being particularly suitable for the study
of signals which are known only inside a limited temporal window. We
like to stress that wavelet transform is not intended to replace the
Fourier transform, which remains very appropriate in the study of
all topics where there is no need for local information.
Quantitatively, given the integer scale parameter $m$, the discrete
wavelet transform (DWT) of a signal $f(t)$ is defined as
\begin{equation}\label{delta1}
W_f (n,m)=\sum_t f(t) \psi_{m,n}^* (t),
\end{equation}
where $$\psi_{m,n} (t)=2^{-m}\psi(2^m t -n)$$ is the dilated and
translated version of the {\it mother} wavelet $\psi (t)$, $n$
constituting a time index running on a scale dependent grid whose
spacing is chosen according to the uncertainty relations; this
implies that discrete wavelet analysis provides good time resolution
and poor frequency resolution at high frequencies and good frequency
resolution and poor time resolution at low frequencies.
Many mother wavelets have been proposed (Haar, Daubechies, Coiflets,
Symlet, Biorthogonal and etc.): the selection of $\psi$ should be
made carefully to better approximate and capture the transient
dynamics of the original signal. 

The {\it \`a trous} wavelet transform (also called stationary wavelet transform SWT) is a wavelet transform algorithm designed to overcome the lack of translation-invariance of the discrete wavelet transform. Translation-invariance is achieved by removing the downsamplers and upsamplers in the DWT.  The SWT is an inherently redundant scheme as the output of each level of SWT contains the same number of samples as the input – so for a decomposition of N levels there is a redundancy of N in the wavelet coefficients. The input data is decomposed into a set of band-pass  filtered  components,  the  wavelet  coefficients,
plus  a  low-pass  filtered  version  of  the  data,  the
continuum (or background or residual). 
Given   a  signal $x (t)$, the {\it \`a trous} wavelet transform decomposes it as a sum of a smooth version of the signal and several {\it detail} signals which take into account the features of the signal at the various scale. Let $J$ be the maximal scale, then $x$ is written as follows: $$x(t)=c_J(t)+\sum_{j=1}^J w_j(t)$$
where $$c_j(t)=\sum_{n=1}^5 h(n)\; c_{j-1} \left(t-2^{j-1}(n-1)\right),$$
$$w_j(t)=c_{j-1}(t)-c_j (t)$$
and $h={1\over 16}$[1 4 6 4 1] is the B-spline filter.
The signals $c_j$ are coarse or smooth version of the original signal, whilst the wavelet coefficients $w_j$
represents the {\it details} of x at scale $2^{-j}$. The indexing is such that $j=1$ corresponds to the finest scale (high frequencies). The maximum scale $J$ is considered as an input.
Unlike widely used non-redundant   wavelet   transforms, it   retains   the
same computational requirement (linear, as a function  of  the  number  of  input  values).  Redundancy (i.e. each scale having the same number of samples
as  the  original  signal)  is  helpful  for  detecting  fine
features in the detail signals since no aliasing biases arise  through  decimation.  However  this  algorithm is  still  simple  to  implement  and  the  computational
requirement  is O(N) per  scale.

\section{Multiscale Granger causality}
Granger causality is a powerful and widespread data-driven  approach
to determine whether and how two time series exert direct dynamical
influences on each other \cite{hla}. 
Quantitatively, let us consider $n$ time series $\{x_\alpha
(t)\}_{\alpha =1,\ldots,n}$; the lagged state vectors are denoted
\begin{equation}\label{delayvectors}
X_\alpha (t)= \left(x_\alpha (t-m),\ldots,x_\alpha (t-1)\right),
\end{equation}
$m$ being the order of the model (window length). Let $\epsilon
\left(x_\alpha |{\bf X}\right)$ be the mean squared error prediction
of $x_\alpha$ on the basis of all the vectors ${\bf
X}=\{X_\beta\}_{\beta =1}^n$. The multivariate  Granger causality
index $\delta_{mv} (\beta \to \alpha )$ is defined as follows:
consider the prediction of $x_\alpha$ on the basis of all the
variables but $X_\beta$ and the prediction of $x_\alpha$ using all
the variables, then the GC is the (normalized) variation of the
error in the two conditions, i.e.
\begin{equation}\label{mv}
\delta_{mv} (\beta \to \alpha )= \log{\epsilon \left(x_\alpha |{\bf
X}\setminus X_\beta\right)\over \epsilon \left(x_\alpha |{\bf
X}\right)};
\end{equation}
The pairwise GC is given by:
\begin{equation}\label{bv}
\delta_{bv} (\beta \to \alpha )= \log{\epsilon \left(x_\alpha
|X_\alpha \right)\over \epsilon \left(x_\alpha |X_\alpha ,
X_\beta\right)}.
\end{equation}

Here we propose to measure the causality $\beta \to \alpha$, at scale $j$, as 
\begin{equation}\label{multiscale}
\Delta_{j} (\beta \to \alpha )= \log{\epsilon \left(x_\alpha
|X_\alpha \right)\over \epsilon \left(x_\alpha |X_\alpha ,
W_j\right)},
\end{equation}
where $$W_j (t)= \left(w_j (t-m),\ldots,w_j (t-1)\right),$$ 
is the vector of detail coefficients of the candidate driver time series $x_\beta$ at scale j.
In other words,  we substitute the single test where the driver time series is considered as a whole, with multiple testing where a single scale is candidate driver, while obviously correcting for multiple comparison with Bonferroni correction as in \cite{noiprl}. 
As an example, we tested the approach on the simulated two time series
\begin{figure}[h]
   \centering
   \includegraphics[scale=.12]{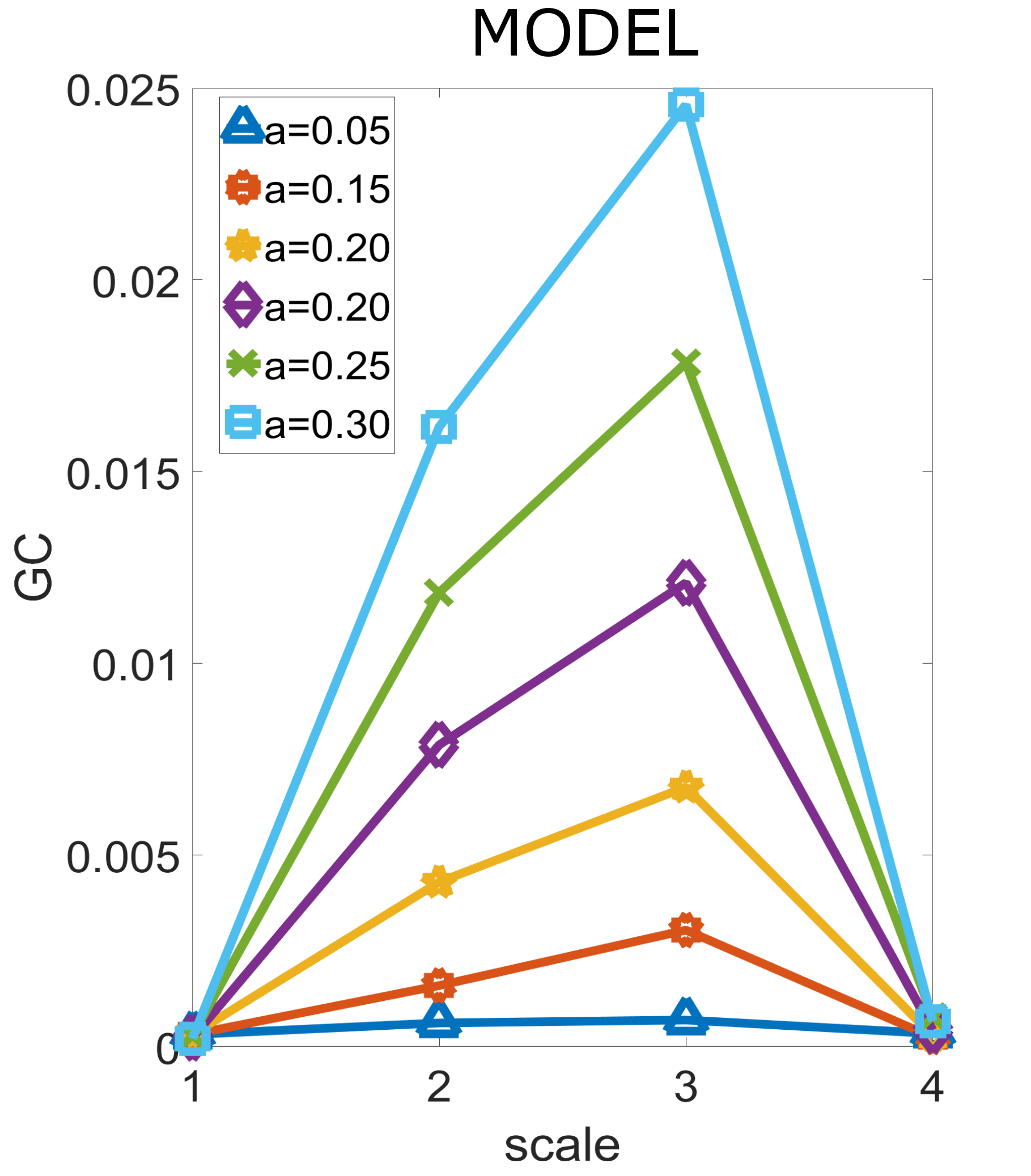}
    \caption{We depict the multiscale GC evaluated by the proposed approach on a simulated two dimensional linear system, unidirectionally coupled with lag equal to 8 and strength a (see the text): it increases with the strength and peaks in correspondence of the lag.}
    \label{fig:model2}
\end{figure}
	 
\begin{eqnarray}
\begin{array}{l}
x (t)= 0.3\; x(t-1) + 0.5\;\eta_1(t)\\
y (t)= 0.1\; y(t-1) + a\;x(t-8)+0.5\;\eta_2 (t)
\end{array}
\label{toy}
\end{eqnarray}
where $\eta_1$ and $\eta_2$ are i.i.d. unit variance noise terms, and $a$ is the coupling $x\to y$. Setting the maximal scale $J=4$, we plot in fig.1 the multiscale GC $\Delta_j$ as evaluated by the proposed approach: it shows a peak in correspondence of the lag, as depicted in figure 1.

\begin{figure}[h]
   \centering
   \includegraphics[scale=.2]{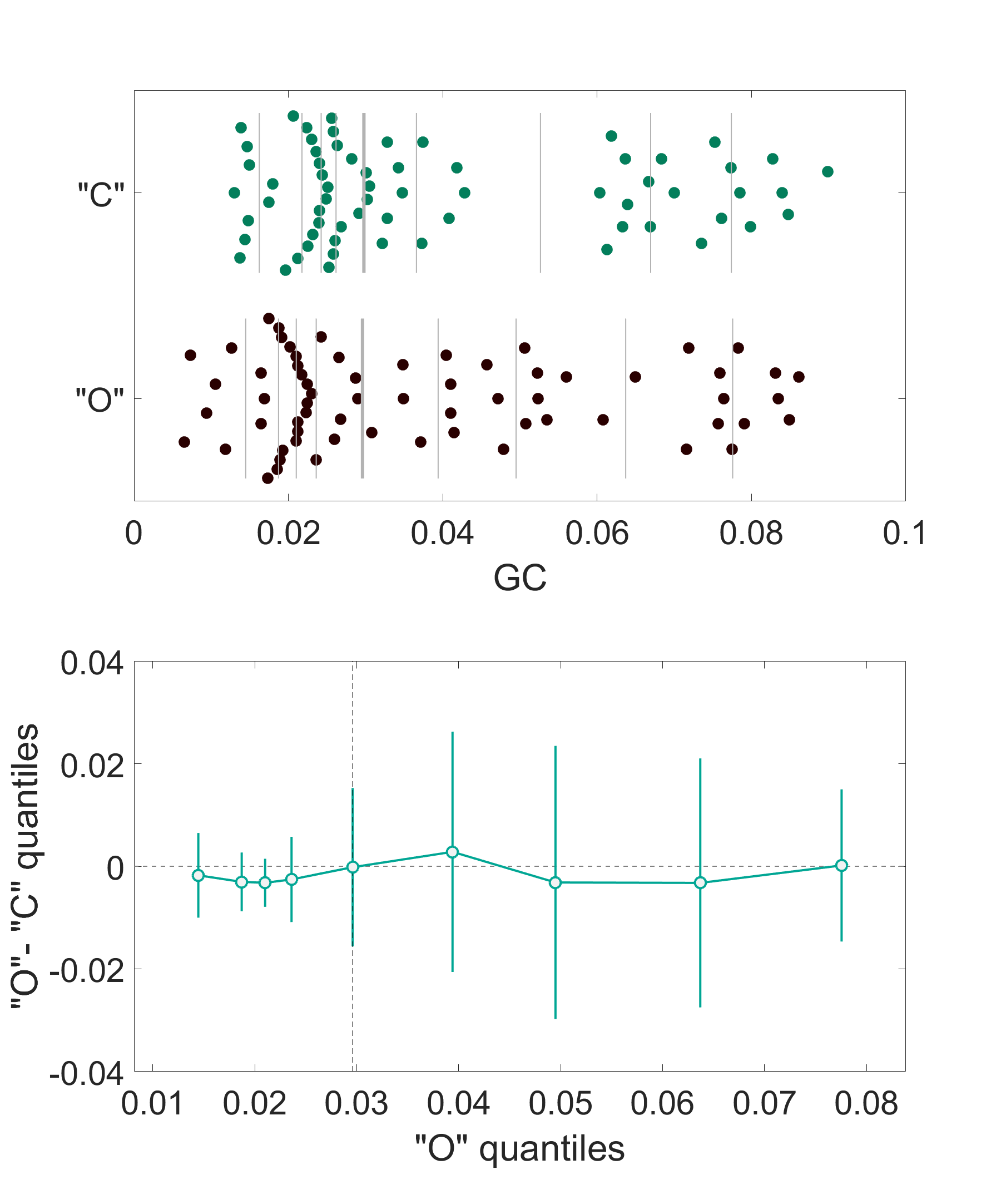}
    \caption{The standard GC, averaged over pairs of channels, is depicted for the 64 subjects in the two conditions, eye open (O) and eye closed (C). In the bottom, the shift function \cite{guillame} is depicted.}
    \label{fig:model}
\end{figure}

\begin{figure}[h]
\centering
\includegraphics[scale=.2]{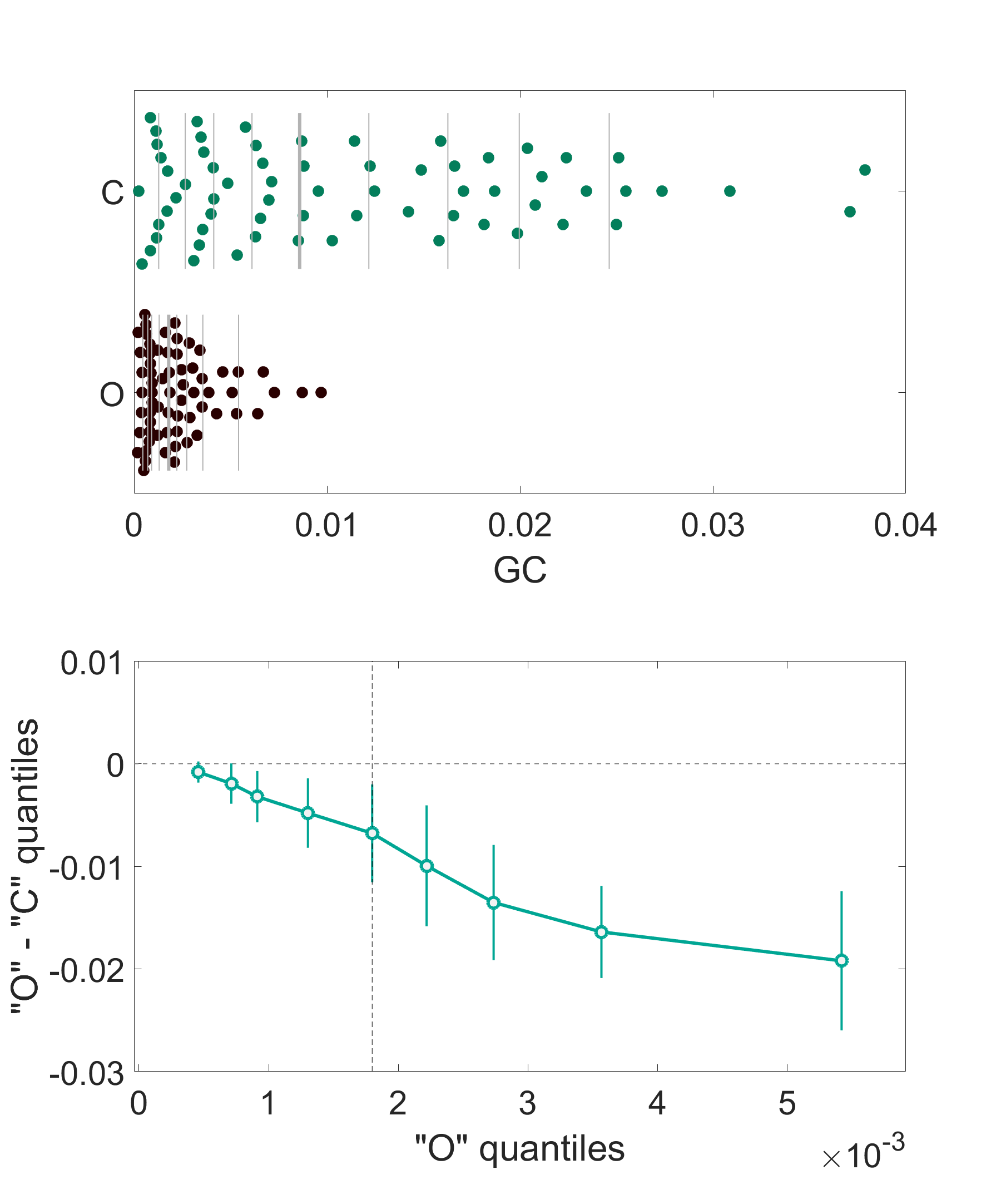}
\caption{The multiscale GC at scale $j=4$, averaged over pairs of channels, is depicted for the 64 subjects in the two conditions, eye open (O) and eye closed (C). In the bottom, the shift function \cite{guillame} is depicted, showing a clear separation in the two conditions for slow scales.}
\label{fig:model1}
\end{figure}

\section{Data set and results}
\label{section4}
We apply the proposed algorithm to scalp EEG signals  gathered   from   the   public  database  PhysioNet  BCI  \cite{physionet}.  The  database  consists  of
healthy  subjects  recorded  in  two  different  baseline
conditions,  i.e. eyes open (EO)  resting  state  and  eyes closed (EC)
resting  state.  In  each  condition,  subjects  were  comfortably
seated  on  a  reclining  chair  in  a  dimly  lit  room.  During  EO
they were asked to avoid ocular blinks in order to reduce signal
contamination. The EEG data were recorded with a 64-channel
system with an original sampling rate of 160 Hz. All the EEG signals are here referenced to the mean signal  gathered  from  electrodes  on  the  ear  lobes. Same data were analyzed in \cite{davico}.
From  64  subjects  and two conditions  (EO,  EC)  the  EEG  signals   epochs  of  10  seconds are considered.  These  epochs  are  considered  as  different  observations  of  the  same  mental  state  and
they  are  used  to assess the differences in directed dynamical connectivity in the two conditions.
We evaluate the multiscale GC here proposed in all the EEG segments and average it over all pairs of channels  in the two conditions for all scales $j=1,\ldots,4$; we also evaluate the classical  GC for all EEG epochs.

We find that the global amount of GC among signals is significantly decreasing as the scale $j$ is increased, in both conditions. Furthermore, comparing signals corresponding to resting conditions with closed eyes and with open eyes, we find that at large scales the directed dynamical connectivity, in terms of the proposed measure, is significantly increased when eyes are closed w.r.t. eyes open, whilst using the standard GC no differences between the two conditions are found.
Standard GC values for eyes open and closed are depicted in figure 2; the multiscale GC from wavelet coefficients (scale 4) is depicted in figure 3. The latter results in a clear separation of the two conditions at all the quantiles.


\section{Conclusions}
\label {conclusion}
A great need exists of effective approaches to measure scale-dependent directed dynamical connectivity, especially in applications where interactions coexist at several scales. Here we have proposed a novel method based on wavelet transform. As Granger causality examines how much the predictability of the target from its past improves when the driver variables’ past values are included in the regression, we measure it at a given scale by including the wavelet coefficients of the driver time series in the regression model of the target. 
Comparing scalp EEG signals in resting subjects we have shown that the wavelet-based multiscale GC at slow scales  significantly increases when eyes are closed (w.r.t. open eye condition); this phenomenon is not detected by the classical GC estimation.

\bibliographystyle{IEEEtran}
\bibliography{bibliography}

\end{document}